\documentclass[preprint,tightenlines,showpacs,amsmath,amssymb]{revtex4}

\usepackage{epsfig,epsf,graphics,psfrag}
\usepackage{times}
\usepackage{float}
\usepackage{color}
\usepackage{amsfonts,amssymb,stmaryrd,latexsym,amsmath}
\usepackage{multirow}
\usepackage{array} 
\usepackage{enumerate}

\newcommand{\psipp}{\psi^\prime\pi\pi}
\newcommand{\hpp}{h_c\pi\pi}
\newcommand{\jppp}{J/\psi(\psi^\prime)\pi\pi}

\newcommand{\jpsipp}{J/\psi\pi\pi}

\begin{document}

\bibliographystyle{unsrt}

\title{Influence of threshold effects induced by charmed meson rescattering}

\author{Xiao-Hai Liu$^1$\footnote{liuxiaohai@pku.edu.cn}}

\affiliation{ Department of Physics and State Key Laboratory of Nuclear Physics and Technology, \\
Peking University, Beijing 100871, China}

\date{\today}

\begin{abstract}
We investigate the processes $e^+e^-$ annihilating to $J/\psi\pi\pi$,
$\psi^\prime\pi\pi$ and $h_c\pi\pi$. The coupled-channel effects induced by the couplings between the widely accepted $D$-wave charmonium
$\psi(4160)$ and $D_1D$, $D_1D^*$ and $D_2D^*$ charmed meson pairs
in the $S$-wave state with couplings given by heavy quark spin symmetry are analyzed. The line shapes show the presence of
cusps that result from the singularities of the
rescattering loops, which could be helpful in understanding the
nature of $Y(4260)$, $Y(4360)$, $Z_c(3900)$/$Z_c(3885)$ and
$Z_c(4020)$/$Z_c(4025)$.


\pacs{~13.25.Gv,~14.40.Pq,~13.75.Lb}
\end{abstract}

\maketitle

\section{Introduction}

There has been a renewal of QCD spectroscopy in the past decade,
initiated by the findings of numerous $XYZ$ states near the
open-flavor thresholds. Most of these states do not fit into the
predictions of the quenched potential quark model, which has been
proved to be very successful in describing the conventional heavy
quarkonia below the open-flavor threshold. These inconsistences remind
people that the vacuum polarization effect of dynamical fermions
should receive more attention in understanding the heavy quarkonium
spectroscopy. This vacuum polarization effect could be described
by the coupled-channel effects induced by the couplings between heavy
quarkonia and open-flavor mesons. After taking into account the
coupled-channel effects, the masses and decay properties of the heavy
quarkonia will be changed significantly, especially when the masses of
the heavy quarkonia are close to the corresponding open-flavor
thresholds~\cite{Eichten:1978tg,Eichten:2005ga,Pennington:2007xr,Barnes:2007xu,Li:2009ad,Guo:2010ak,vanBeveren:2009ka,vanBeveren:2009hb,Zhou:2013ada}.

The mysterious charmonium-like state $Y(4260)$ has many peculiar
properties. As a charmonium candidate, it is observed in the
$\jpsipp$ channel, but not in the open-charm decay channels which
are supposed to be favorable decay modes of conventional $c\bar{c}$
states. The $R$-value scan around 4.26 GeV also appears to have a
dip instead of a bump structure. The state observed in the $\psipp$
channel, $Y(4360)$, has similar puzzles as those of $Y(4260)$.
Recent experimental observations revive discussions on the nature of
$Y(4260)$. Several charged charmonium-like structures, $Z_c(3900)$,
$Z_c(3885)$, $Z_c(4020)$ and $Z_c(4025)$, are observed while
studying $Y(4260)$
\cite{Ablikim:2013mio,Liu:2013dau,Ablikim:2013wzq,Ablikim:2013emm,Ablikim:2013xfr,Xiao:2013iha},
which makes $Y(4260)$ more intriguing. We refer to
Refs.~\cite{Liu:2013waa,Olsen:2014mea} for a recent review about these $XYZ$ states.

Since the masses of excited charmed mesons are usually larger, the influence of coupled-channel effects on charmonia with the
$P$-wave charmed mesons ($D_0$, $D_1$, etc.) involved has not been widely studied before. On the other hand, the thresholds
of the combinations of $S$- and $P$-wave charmed mesons are very
close to $Y(4260)$ and $Y(4360)$, and their couplings with the
parity-odd charmonia could be $S$ wave, which is supposed to be
strong. In this paper, we will study the influence of the
coupled-channel effects on the line shapes of some pertinent cross
sections and invariant mass distributions, where the contributions
with $P$-wave charmed mesons involved are emphasized.

\section{Coupled-channel effects in the dipion transitions}

We will build our model within the framework of heavy hadron chiral
perturbation theory (HHChPT). In HHChPT, to encode the heavy quark
spin symmetry (HQSS), the doublets with light degrees of freedom (LDOF)
$J^P=1/2^-$, $1/2^+$, $3/2^+$ are collected into three superfields
\begin{eqnarray}
 H_a  &=& \frac{1+{\rlap{v}/}}{2}[\mathcal{D}_{a\mu}^*\gamma^\mu-\mathcal{D}_a\gamma_5] \ , \\
  S_a &=& \frac{1+{\rlap{v}/}}{2} \left[\mathcal{D}_{1a}^{\prime \mu}\gamma_\mu\gamma_5-\mathcal{D}_{0a}^*\right] \ ,\\
   T_a^\mu &=&\frac{1+{\rlap{v}/}}{2} \bigg\{ \mathcal{D}^{\mu\nu}_{2a} \gamma_\nu \nonumber \\
    &-&\sqrt{3 \over 2}\mathcal{D}_{1a\nu}  \gamma_5 \left[
g^{\mu \nu}-{1 \over 3} \gamma^\nu (\gamma^\mu-v^\mu) \right]
\bigg\} \ ,
\end{eqnarray}
respectively, where $a$ is the light flavor index. For clarity, in
the following, we will use $HH$ to represent $D^{(*)}D^{(*)}$
combinations, with the similar conventions for $TH$ and $SH$. The
$S$-wave charmonia $\psi(nS)$ may couple to $HH$ and $SH$ via
relative $P$ and $S$ wave respectively, where $n$ denotes the
radial quantum number. Required by HQSS, the total angular momentum
of LDOF should also be conserved for these
couplings. For the $TH$ combination, their LDOF carry
angular momentum $3/2$ and $1/2$ respectively. In an $S$-wave coupling, they
cannot produce zero angular momentum carried by LDOF of $\psi(nS)$. As a result, the $S$-wave coupling between
$\psi(nS)$ and $TH$, although allowed by the parity conservation,
will be suppressed according to HQSS~\cite{Li:2013yka}. In the heavy quark limit, the only allowed
coupling is $D$ wave. However, for the coupling between $D$-wave
charmonia $\psi(nD)$ and $TH$, the $S$-wave coupling is allowed, since the total angular momentum of the
LDOF of $\psi(nD)$ is 2. This gives us a hint that the
coupled-channel interactions between $\psi(nD)$ and $TH$ may largely
affect the mass and decay properties of $\psi(nD)$, especially for
the $D$-wave charmonia close to $TH$ thresholds. In the charmonium
family, $\psi(4160)$ is widely considered as a conventional $2^3D_1$
charmonium, whose mass and width are estimated by PDG to be
$M_{\psi}$=$4153\pm 3$ MeV and $\Gamma_{\psi}$=$103\pm 8$
MeV~\cite{Beringer:1900zz}. If we use the latest data, its mass and
width are found to be $M_{\psi}$=$4191.7\pm 6.5$ MeV and
$\Gamma_{\psi}$=$71.8\pm 12.3$ MeV in Ref.~\cite{Ablikim:2007gd}, or
$M_{\psi}$=$4193\pm 7$ MeV and $\Gamma_{\psi}$=$79\pm 14$ MeV in
Ref.~\cite{Mo:2010bw}, which are very close to $TH$ thresholds
and the mass of $Y(4260)$. Therefore it is natural to wonder whether
there is some kind of connection between these states and the $TH$
coupled channels. For further discussion, we mention that $HH$ may
couple to $\psi(nD)$ via the $P$ wave, and the $S$-wave coupling between
$SH$ and $\psi(nD)$ is also suppressed according to HQSS. $SH$ may
couple to $\psi(nS)$ via the $S$ wave. We show the HQSS allowed couplings in Table.~\ref{hqss} to make the above points clear.

\begin{table}[htbp]
  \caption{HQSS allowed couplings.\label{hqss}}
\begin{center}
\begin{tabular}{|c|c|c|c|}
  \hline
   & $HH$ & $SH$ & $TH$ \\
   \hline
  $\psi(nS)$ & $P$-wave & $S$-wave & $D$-wave \\
  \hline
  $\psi(nD)$ & $P$-wave & $D$-wave & $S$-, $D$-wave \\
  \hline
\end{tabular}
\end{center}
\end{table}

There are plenty of data for $e^+e^-$ annihilating to one heavy
quarkonium plus two pion mesons. Many interesting phenomena have
been discovered in these channels. We will investigate these
exclusive processes in this paper to try to quantify the
coupled-channel effects. Taking into account the previous
discussions, using HHChPT power counting we introduce the leading
order effective Lagrangian as follows:
\begin{eqnarray}\label{lag1}
  \mathcal{L}_1 &=& \frac{g_T}{\sqrt{2}} < J^{\mu\nu} \bar{H}^\dag_a \gamma_\nu \bar{T}_{a\mu}  -  J^{\mu\nu} \bar{T}^\dag_{a\mu} \gamma_\nu \bar{H}_a >  \nonumber \\
&+& ig_H < J^{\mu\nu} \bar{H}^\dag_a \gamma_\mu \overleftrightarrow{\partial}_\nu \bar{H}_a >  \nonumber \\
 &+& g_S < J \bar{S}^\dag_a  \bar{H}_a  +  J \bar{H}^\dag_a  \bar{S}_a >  \nonumber \\
 &+& C_S < J \bar{H}^\dag_b \gamma_\mu\gamma_5 \bar{H}_a \mathcal{A}^\mu_{ba} >  \nonumber \\
 &+& iC_P < J^\mu \bar{H}^\dag_b \sigma_{\mu\nu}\gamma_5 \bar{H}_a \mathcal{A}^\nu_{ba} > +H.c.,
\end{eqnarray}
where $<\cdot\cdot\cdot>$ means the trace over Dirac matrices,
$\mathcal{A}^\mu$ is the chiral axial vector containing the
Goldstone bosons, and the fields for the $S$-, $P$-, and $D$-wave
charmonia read
\begin{eqnarray}
J &=& \frac{1+{\rlap{v}/}}{2} [\psi(nS)^\mu \gamma_\mu]
\frac{1-{\rlap{v}/}}{2}, J^\mu =\frac{1+{\rlap{v}/}}{2}
[h_c(nP)^\mu \gamma_5] \frac{1-{\rlap{v}/}}{2} \nonumber \\
&& J^{\mu\nu} = \frac{1+{\rlap{v}/}}{2} \bigg\{\psi(nD)_\alpha
\bigg[ \frac{1}{2}\sqrt{\frac{3}{5}} [(\gamma^\mu-v^\mu)
g^{\alpha\nu} \nonumber \\
&&+ (\gamma^\nu-v^\nu)g^{\alpha\mu}] -\sqrt{\frac{1}{15}}
(g^{\mu\nu}-v^\mu v^\nu)\gamma^\alpha \bigg] \bigg\} \
\frac{1-{\rlap{v}/}}{2},
\end{eqnarray}
respectively, where only the states relevant for our discussion are
included. The effective Lagrangian for the strong interactions of
heavy mesons with Goldstone bosons reads
\begin{eqnarray}
{\mathcal L}_2 &=&  i{h^\prime \over \Lambda_\chi} < {\bar H}_a
T^\mu_b \gamma^\nu\gamma_5 ( D_\mu \mathcal{A}_\nu + D_\nu
\mathcal{A}_\mu)_{ba} > \nonumber\\
&+& ih < {\bar H}_a S_b \gamma_\mu \gamma_5 {\cal A}_{ba}^\mu > +ig
<H_b \gamma_\mu\gamma_5 \mathcal{A}^\mu_{ba} \bar{H}_a>.
\end{eqnarray}
Some of these Lagrangians have been introduced in
Refs.~\cite{Casalbuoni:1992dx,Casalbuoni:1996pg,Colangelo:2003sa,Mehen:2011yh,Margaryan:2013tta,Guo:2013zbw},
we refer to Ref.~\cite{Casalbuoni:1996pg} for a review and some
conventions.

The coefficient $g_T$ in Eq.~(\ref{lag1}), which describes the
coupling strength between $\psi(nD)$ and $TH$, is not well
determined. But taking into account the coupling is $S$ wave, it may
be expected to be large. There are some indirect experimental
evidences to support this argument. For instance, $\psi(4415)$ is a
widely accepted $S$-wave charmonium, its decaying to $D_2D$ is a
$D$-wave decay, but the branching fraction is very large, which is
$(10\pm 4)\%$ estimated by PDG~\cite{Beringer:1900zz}. Therefore it
seems to be reasonable to expect the $S$-wave coupling constant
$g_T$ could also be sizable. Of course this is not a serious
estimation, to obtain some less model-dependent result, we will only
focus on the line shape behavior of the total and differential cross
sections of the pertinent channels in this paper.

\begin{figure}[htb]
\centering
\includegraphics[width=0.8\hsize]{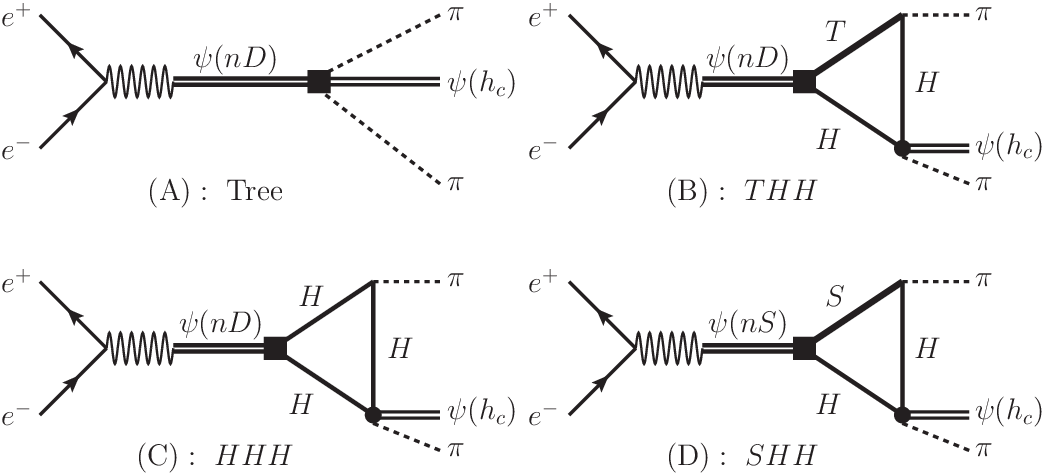}
\caption{Feynman diagrams for the dipion
transitions.}\label{feynmandiag}
\end{figure}

 The Feynman diagrams which contribute to the dipion
transitions are displayed in Fig.~\ref{feynmandiag}, where we will
take $\psi(4160)$ as the most relevant $\psi(nD)$ state, and use its
PDG averaged mass and width as the input parameters in our
calculation. Although the production of $D$-wave charmonia in
$e^+e^-$ annihilation is supposed to be suppressed, the experimental
data indicates the electron decay width of $\psi(4160)$ is not
small, i.e. $\Gamma_{ee}$=$(0.83\pm 0.07)$ KeV
~\cite{Beringer:1900zz}, which possibly results from the $S$-$D$ mixing
effect. However, we still face a dilemma here. The larger electron
decay width of $\psi(4160)$ implies the HQSS breaking effects, or
some higher order contributions in the effective theory, may also be
important, which can be ascribed to the fact that the charm quark is not so heavy.
For the moment we will ignore some symmetry breaking effects, such
as the breaking in the couplings between $\psi(nD)$ and $TH$, and
still follow the guidance of HQSS (see
Refs.~\cite{Li:2013yka,Wang:2013kra} for some discussions on the
symmetry breaking effects).

For the tree diagram displayed in Fig.~\ref{feynmandiag}(A),
its amplitude will be proportional to a Breit-Wigner form
function
\begin{eqnarray}\label{bw4160}
BW[\psi(4160)]=(s-M_\psi^2+iM_\psi \Gamma_\psi)^{-1},
\end{eqnarray}
where $s$ is the center of mass energy squared. The cross section
line shape of this diagram will be ordinary, which is just the usual
Breit-Wigner structure, but it can provide some background that may
affect the line shape behavior via interference.

\begin{figure}[htb]
\centering
\includegraphics[width=0.3\hsize]{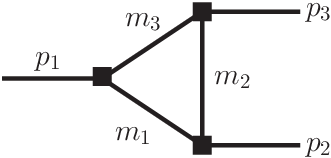}
\caption{Triangle diagram under discussion. For the external momenta, we define $p_1^2=W^2$, $p_2^2=s_2$, and $p_3^2=m^2$.}\label{triangle}
\end{figure}

For the triangle diagrams
which describe the coupled-channel effects, there are several kinds of singularities corresponding to them.
The location of the singularities in the complex space of the external momentum variables can be determined by a set of equations, which are usually called the Landau equations~\cite{Landau:1959fi}. In some special
kinematic configurations, all of the three internal lines can
be on shell simultaneously, which corresponds to the \textit{leading singularity} of the triangle diagram~\cite{Eden:1966}. The singularities that correspond to two of the internal lines being on shell are \textit{lower-order singularities}~\cite{Eden:1966}. For the triangle diagram, the location of the \textit{leading singularity} corresponds to the \textit{anomalous threshold}, while the \textit{lower-order singularity} corresponds to the \textit{normal threshold} \cite{Eden:1966,Landshoff:1962,Landshoff:1961}. For instance, in Fig.~\ref{triangle}, when $W$$=$$m_1$$+$$m_3$, the \textit{anomalous threshold} $\bar{s}_2$ in the complex $s_2$-plane is real, and
\begin{equation}\label{threshold}
\bar{s}_2=s_n +\frac{m_1}{m_3} [ (m_2-m_3)^2-m^2 ],
\end{equation}
where $s_n$ is the \textit{normal threshold} $(m_1+m_2)^2$, and we have assumed the internal particles are stable.
The above triangle singularities (TS) are usually branch points of the amplitude in the complex space. When the singularities approach close to the physical region, they may affect the threshold behavior of the physical amplitude dramatically, or show
up directly as bumps or cusps in the amplitude \cite{Landshoff:1962,Eden:1966,Wu:2011yx,Wang:2013cya,Liu:2013vfa}. The $THH$ loops in our discussion just
approximately meet the kinematic conditions of the leading TS, and for the charmed meson loops, according to Eq. (\ref{threshold}), the \textit{anomalous threshold} and \textit{normal threshold} are very close to each other. We therefore expect these TS may lead to some detectable effects for the relevant processes.

\begin{figure}[htbp]
\centering
\includegraphics[width=0.5\hsize]{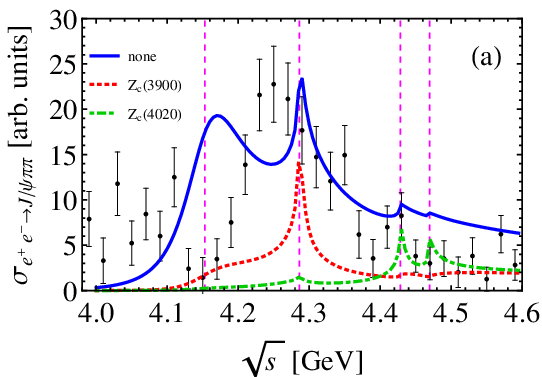}\includegraphics[width=0.49\hsize]{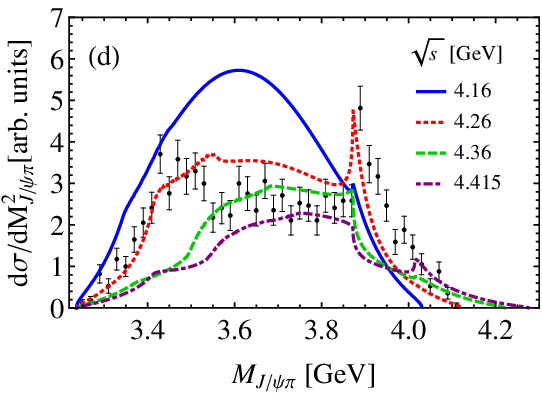} \\
\includegraphics[width=0.5\hsize]{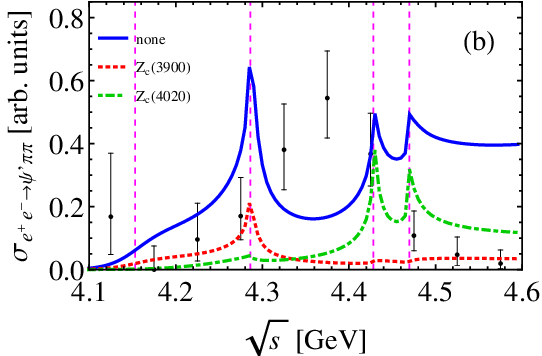} \includegraphics[width=0.49\hsize]{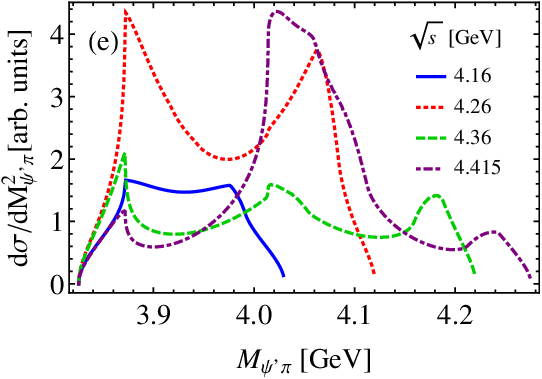}\\
 \includegraphics[width=0.5\hsize]{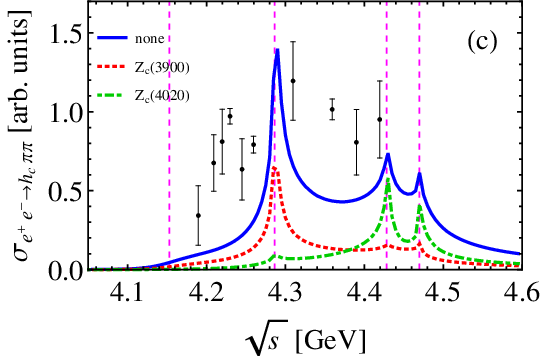}\includegraphics[width=0.5\hsize]{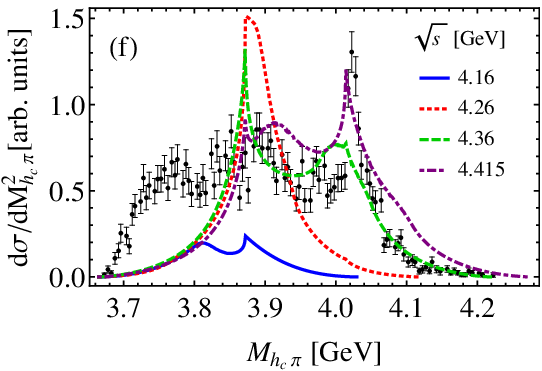}
\caption{\textbf{Left}: Energy dependence of the cross section for
$e^+e^-$$\to$ (a) $J/\psi\pi\pi$, (b) $\psi^\prime\pi\pi$, and (c)
$h_c\pi\pi$ via $\psi(4160)$ and intermediate $THH$ loops. The solid
line is the result with only taking into account the contact
interaction, the dotted and dotdashed lines correspond to the result
with plugging into $Z_c(3900)$ and $Z_c(4020)$ propagator
respectively, and the magnitude of these lines has been rescaled to
a similar level. The vertical lines (from left to right) indicate
the mass of $\psi(4160)$, the thresholds of $D_1D$, $D_1D^*$ and
$D_2D^*$ respectively. \textbf{Right}: The corresponding invariant
mass distributions of (d) $J/\psi\pi$, (e) $\psi^\prime\pi$, and (f)
$h_c\pi$ at four center of mass energy points. Only the contact
interactions are taken into account in $THH$ loops. The experimental data are taken from Ref.~\cite{Yuan:2007sj} for (a), Ref.~\cite{Wang:2007ea} for (b), Ref.~\cite{Ablikim:2013wzq} for (c), Ref.~\cite{Ablikim:2013mio} for (d), and Ref.~\cite{Ablikim:2013wzq} for (f) respectively. The data points are rescaled to adapt the theoretical lines. }\label{totxs}
\end{figure}

Concerning Fig.~\ref{feynmandiag}(B), named as $THH$ loop in this
paper, there are four subdiagrams categorized by the intermediated
charmed mesons:
\begin{enumerate}[I)~]
  \item  $\{ D_1 D\ [D^*] \}$,
  \item  $\{ D_1 D^* \ [D^*] \}$,
  \item  $\{ D_2 D^* \ [D] \}$,
  \item  $\{ D_2 D^* \ [D^*] \}$,
\end{enumerate}
where the charmed mesons in the brackets correspond to the vertical
propagators in the $THH$ loops.
For $\jppp$ final states, the amplitudes corresponding
to the above four subdiagrams have a simple relation in the heavy
quark limit, i.e.,
\begin{equation}\label{ampratio}
\mathcal{M}^{I}:\mathcal{M}^{II}:\mathcal{M}^{III}:\mathcal{M}^{IV}=1:\frac{1}{2}:-\frac{1}{5}:\frac{3}{10}
\ .
\end{equation}
This implies the main contribution may come from the $\{ D_1 D\ [D^*]
\}$ loop. For $\psi(nD)$$\to$$h_c\pi\pi$, the spin of the charm
quark is flipped, which means this process
is forbidden in the heavy quark limit. However, since we are using physical masses as input in
the calculation and the masses of charmonia and charmed mesons
are not so heavy, the amplitude still could be sizable. In
Fig.~\ref{totxs}, we display the line shapes of the energy dependence
of the cross sections for $e^+e^-$$\to$$\jpsipp$, $\psipp$ and
$\hpp$ via the $D$-wave state $\psi(4160)$ and $THH$ loops. For the
$\jpsipp$ channel, apart from the $\psi(4160)$ bump, three cusps
appeared at the thresholds of $D_1D$, $D_1D^*$ and $D_2D^*$
respectively. Among these cusps, the one staying around $D_1D$ threshold is the most obvious one, which can be
understood according to Eq.~(\ref{ampratio}). The peak position of
$\psi(4160)$ is upward shifted because of the interference with
$D_1D$ cusp. For $\psipp$ and $\hpp$ channels, the $\psi(4160)$ bump
is nearly smeared since the $D_1D$ cusp is much stronger. This can
be attributed to the thresholds of $\psi^\prime\pi$ and $h_c\pi$ are
much closer to that of $HH$, compared with that of $J/\psi\pi$,
which will strengthen TS. There is another way to understand
this point. If we assume there are $Z_c(3900)$ and $Z_c(4020)$
molecular states produced in this $THH$ loop mechanism, which
corresponds to plug two propagators into the black bubble of the
diagram Fig.~\ref{feynmandiag}(B) separately, the line shapes will be
changed to some extent, as illustrated in Fig.~\ref{totxs}. Taking
into account the four types of $THH$ loops, if we plug in
$Z_c(3900)$, since it stays in the vicinity of $DD^*$ threshold,
$\mathcal{M}^{I}$ will become so strong that the $D_1D^*$ and
$D_2D^*$ cusps blur to obscurity. On the other hand, if we plug in
$Z_c(4020)$ which is much closer to the $D^*D^*$ threshold,
$\mathcal{M}^{II}$ and $\mathcal{M}^{IV}$ will be strengthened,
therefore the $D_1D^*$ and $D_2D^*$ cusps become more obvious.

The line shapes of differential cross sections also show some
extraordinary phenomena. We display the results at several center of
mass energy points in Fig.~\ref{totxs}. For
$J/\psi\pi$ distribution, a mini cusp appeared around the
$DD^*$ threshold at $\sqrt{s}=$$4.16$ GeV. With the energy
increasing and being close to the $D_1D$ threshold, TS may occur and
a clear narrow cusp emerged around the $DD^*$ threshold at
$\sqrt{s}=$$4.26$ GeV. With the continuous growth of the energy, the
contributions of $\mathcal{M}^{II}$, $\mathcal{M}^{III}$ and
$\mathcal{M}^{IV}$ become more and more significant, and the cusp
around $D^*D^*$ is emerging, but the cusp around $DD^*$ is fading
since the energy is running away from the favorable region where TS
of $\mathcal{M}^{I}$ plays an important role. For $\psi^\prime\pi$ and $h_c\pi$
distributions, the $D^*D^*$ cusp has already showed up around
$\sqrt{s}=$$4.36$ GeV. When the energy comes to $4.415$ GeV, the
$D^*D^*$ cusp becomes very obvious. It should be mentioned these
cusps are also affected by the reflection effects in the Dalitz
plot.

From the above discussions, it can be concluded that the kinematics
play a crucial role to produce the intriguing line shapes of total and
differential cross sections. That is because TS of the $THH$
loops has its favorable kinematic region, it is sensitive both to
the masses of the internal and external states. Another important
relevant factor is the phase space. These factors lead to the
production of different cusps for different final states and
different energy points. The $\psi^\prime\pi\pi$ channel can be
taken as a direct prediction to check this argument, since its
dynamical production mechanism is the same with that of $\jpsipp$
but the kinematics is different.

If comparing Fig.~\ref{totxs} with the experimental data in Refs.~\cite{Ablikim:2013mio,Liu:2013dau,Ablikim:2013wzq,Xiao:2013iha,Yuan:2013ffw,Yuan:2007sj,Wang:2007ea,Aubert:2008aj,Aubert:2007zz},
it can be noticed that these cusps approximately fall in the corresponding
vicinities of $Y(4260)$, $Y(4360)$, $Z_c(3900)$/$Z_c(3885)$ and
$Z_c(4020)$/$Z_c(4025)$ in the same processes, and there are no
genuine resonances introduced in this model. These cusps are
generated in the dipion transitions by this special rescattering
mechanism, but the open-charm channels
$D^{(*)}_{(s)}D^{(*)}_{(s)}$ will not suffer from this mechanism.
Therefore it will not be very surprising to observe a dip in
$R$-value scan and open charm distributions around $4.26$
and $4.36$ GeV. Apart from this, there is $DD^*$ ($Z_c(3900)$)
but no $D^*D^*$($Z_c(4020)$) threshold bump obtained in the $J/\psi\pi$
distribution, which is in agreement with the experimental
observations~\cite{Ablikim:2013mio,Liu:2013dau}. For the $h_c\pi\pi$ channel, there is a distinct $Z_c(4020)$ signal but no significant $Z_c(3900)$ signal observed in the experiment~\cite{Ablikim:2013wzq}. On one hand, in our $THH$ loop mechanism, the line shape behavior of the differential cross section is sensitive to the kinematics, and different cusps will appear for different center of mass energy $\sqrt{s}$. For instance, as illustrated in Fig.~\ref{totxs}(f), when $\sqrt{s}$=4.26 GeV, the $DD^*$ cusp is much more obvious than the $D^*D^*$ cusp. In contrast, when $\sqrt{s}$=4.415 GeV, the $D^*D^*$ cusp is more obvious. On the other hand, the experimental data displayed in Fig.~\ref{totxs}(f) is a summation over data at many energy points, and the integrated luminosities and cross sections are different among these energy points~\cite{Ablikim:2013wzq}. Considering that in our model the
relative strength of $DD^*$ and $D^*D^*$ cusps will change according to the center of mass energy, we qualitatively suppose that the $THH$ loop mechanism can partly account for the experimental observation of the $h_c\pi\pi$ channel. It should be mentioned that, in Fig.~\ref{totxs}, we incorporate some data points of the pertinent experiments, but we do not mean to fit the data considering these plots only include the contributions from $THH$ loops.

However, just according to this simple model, the peak positions and
bump structures are not precisely consistent with the current data.
For instance, there is still a shifted $\psi(4160)$ bump appearing in
Fig.~\ref{totxs}(a), but this structure is not clear in
experiment~\cite{Yuan:2007sj}. Another inconsistency is that $DD^*$
cusps are stronger than $D^*D^*$ cusps at $\sqrt{s}=$$4.26$ and
$4.36$ GeV in Fig.~\ref{totxs}(f) (compared with
Ref.~\cite{Ablikim:2013wzq}), although we concluded that the
strength of the cusps are sensitive to energies. To compensate for
the deficiency of this simplified scenario, it seems that we need
some proper interferences between the tree diagrams and $THH$
loops. This seems to be possible, since the tree diagrams will
only affect the energy region around $M_{\psi(4160)}$ according to
the Breit-Wigner function Eq.~(\ref{bw4160}), a proper destructive
interference will possibly flatten out the bump around $\psi(4160)$.
On the other hand, for $\psipp$ and $\hpp$ final states, as the
$\psi(4160)$ structure is already smeared, the interference may
possibly make it show up again and change the $D_1D$ cusp structure.
With the center of mass energy increasing, the contribution from
some other higher charmonia, such as $\psi(4415)$, will be involved.
$\psi(4415)$ also stays close to the thresholds of $TH$, but these
couplings are $D$ wave, whose contribution will be higher order and
the cusps are expected to be weakened. If we take into account $S$-$D$
mixing between charmonia, $\psi(4415)$ can also couple to $TH$ via
$S$ wave. Since its mass is closer to $D_1D^*$ compared with
$D_1D$, it will strengthen $\mathcal{M}^{II}$ and then strengthen
the $D^*D^*$ cusp. This may also compensate for the deficiency
appearing in the $h_c\pi\pi$ channel. However, since we cannot give
reliable estimations of the pertinent couplings for the moment,
these are just some qualitative speculations.

There are also some theoretical uncertainties concerning our
scenario. For $HH$$\to$$J/\psi(\psi^\prime)\pi$, there are two
mechanisms that contribute at the same order according to the HHChPT
power counting. One is the short distance process mediated by the
contact interaction, as we used in our model. Another one is the
$t$-channel process by exchanging an off shell charmed meson. For
$HH$$\to$$h_c\pi$, the contribution from the second one is even at a
lower order. If we take the $t$-channel interaction into account, it
will change the triangle diagram to the box diagram. But the
singular properties of the box diagram can be ascribed to the
triangle diagrams, and the most important contribution still comes
from the case when $THH$ are approximately on shell. To simplify the
calculation and show the intrinsic characters of the loops,
i.e. TS, we will mainly discuss the triangle diagrams here. The
relative strength of the rescattering amplitude will be affected by
these theoretical uncertainties, but the singularity behavior of the
loops is mainly in connection with the kinematics, and the
line shapes will not be distorted much.

From the point of view of TS and kinematics, the model discussed
here shares the similar scenario with the $D_1 D$ molecular state
ansatz discussed in
Refs.~\cite{Wang:2013cya,Liu:2013vfa,Cleven:2013mka,Wang:2013hga}. One different
point is that it incorporates the $D_1D$, $D_1D^*$ and $D_2D^*$
combinations in a singe Lagrangian with the relative phase and
coupling strength fixed in the heavy quark limit, which leads to
most of the $TH$ and $HH$ cusp structures being studied
simultaneously in the same channel. Another crucial point is, no
matter whether the molecular state exists or not, it seems to be
natural to suppose the coupled-channel effects, or the vacuum
polarization effects, should exist physically.

\begin{figure}[htbp]
\centering
\includegraphics[width=0.5\hsize]{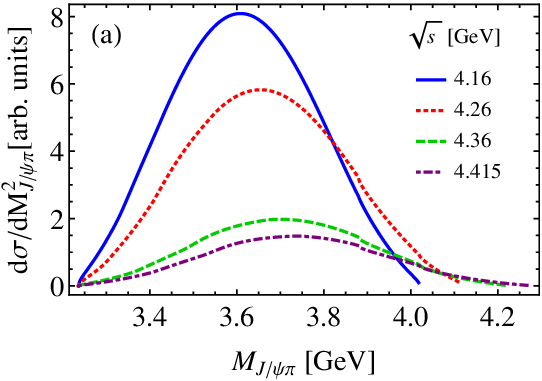}\includegraphics[width=0.52\hsize]{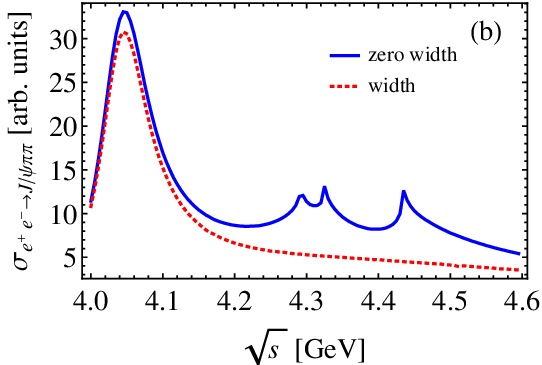}
\caption{(a)Differential cross section for
$e^+e^-$$\to$$J/\psi\pi\pi$ via $\psi(4160)$ and intermediate $HHH$
loops. (b) Energy dependence of the cross section for $e^+e^-$$\to$
$J/\psi\pi\pi$ via $\psi(4040)$ and intermediate $SHH$ loops, where
the solid and dotted line corresponds to the result with and without
taking into account the broad width influence of $D_0$ and
$D_1^\prime$ respectively.}\label{difxsHHH}
\end{figure}

This is not the whole story concerning the rescattering processes if
we only take into account the $THH$ loops. Since experiments
indicate the main decay channel of $\psi(4160)$ is $HH$, we should
also include the contribution from $HHH$ loops, as illustrated in
Fig.~\ref{feynmandiag}(C). But the threshold of $HH$ is far away
from the energy region discussed here, which does not favor the
kinematic conditions of TS, and the coupling between $\psi(nD)$ and
$HH$ is $P$ wave, which will also suppress the rescattering
amplitude. To make it clear, we display the differential cross
section for $e^+e^-$$\to$$\jpsipp$ in Fig.~\ref{difxsHHH}(a). At the
center of mass energy around $\sqrt{s}=4.16 $ GeV, if the
rescattering occurs via $THH$ loops, although it is not the
favorable energy point for producing singularity, there is still a
visible narrow cusp appearing around the $DD^*$ threshold. But if the
rescattering occurs via $HHH$ loops, the cusp will be nearly smeared, which is inconsistent with the result
of CLEO-c~\cite{Xiao:2013iha}. This in another way supports that the
coupling between $\psi(4160)$ and $TH$ could be sizable. We also
studied the rescattering processes through $\psi(nS)$ and $SHH$
loops. As discussed in Ref.~\cite{Liu:2013vfa}, since $D_0$ and
$D_1^\prime$ are too broad, if taking into account their width
effects, the singularities will be smoothed out and the amplitude will
be lowered to some extent. We display one result in
Fig.~\ref{difxsHHH}(b), where we have chosen $\psi(4040)$ as the
intermediate $S$-wave charmonium [another option is $\psi(4415)$]. It
can be seen the cusps at $D_0D^*$, $D_1^\prime D$, and $D_1^\prime
D^*$ are smoothed by the broad width. This is just a simple
estimation, since we only change the propagators in the loops to the
Breit-Winger functions, which may account for the contributions from
higher order corrections. The real situation may be complicated.
Although the line shape behavior of $HHH$ and $SHH$ loops seems to be
ordinary, they can also provide some background for interference with
$THH$ loops, which is similar to the tree diagrams.

\section{Summary}

In conclusion, we have discussed the line shape behavior of the cross
sections and distributions of $e^+e^-$$\to$$\jpsipp$, $\psipp$ and
$\hpp$. The coupled-channel effects, especially that induced by the
couplings between the $D$-wave charmonia and $TH$ charmed mesons
($THH$ loops), are emphasized. Because these leading order $S$-wave
couplings will respect HQSS, and another important reason is the
thresholds of $TH$ are close to that of $Y(4260)$ and $Y(4360)$.
Using $\psi(4160)$ as the input $\psi(nD)$, we obtain some
cusps staying at the thresholds of $TH$ and
$HH$, which may have some underlying connections with the $XYZ$
states observed around these thresholds. With a few theoretical
uncertainties, the line shape behavior is less model dependent, and
it indicates that these cusps are sensitive to the kinematics, that
is because TS of the $THH$ loops has its favorable kinematic
region. This can explain why $Z_c(3900)$/$Z_c(3885)$ and
$Z_c(4020)$/$Z_c(4025)$ are observed in different processes and
energy points. The $\psi^\prime\pi\pi$ channel can be taken as a
direct prediction to check this scenario.

Our paper just focuses on the dipion transitions of the charmonia, as a
qualitative guess, if the coupled-channel effects with the
$P$-wave charmed mesons involved are truly so important, maybe they can also
compensate for the mass shift of charmonia sizably.

\subsection*{Acknowledgments}
The author would like to thank S.L. Zhu, Q. Zhao, F.K. Guo and G. Li
for helpful discussions. This work was supported in part by the
China Postdoctoral Science Foundation under Grant No. 2013M530461,
the National Natural Science Foundation of China under Grants No.
11075004 and No. 11021092, and the DFG and the NSFC through funds
provided to the Sino-German CRC 110 "Symmetries and the Emergence of
Structure in QCD."

\end{document}